\documentclass[pra,aps,showpacs,twocolumn,twoside]{revtex4}

\usepackage{amsfonts}
\usepackage{amsmath}
\usepackage{amssymb}
\usepackage{graphicx}
\usepackage{psfrag}
\usepackage{color}

\definecolor{labelkey}{gray}{.1}
\definecolor{refkey}{gray}{.1}
\definecolor{dgreen}{rgb}{0.000,0.500,0.000}
\begin{document}
\title{Classification of multipartite entangled states by multidimensional determinants}
\author{Akimasa Miyake}
\email{miyake@monet.phys.s.u-tokyo.ac.jp}
\affiliation{ Department of Physics, Graduate School of Science, 
University of Tokyo,\\  Hongo 7-3-1, Bunkyo-ku, Tokyo 113-0033, Japan}
\affiliation{ Quantum Computation and Information Project, ERATO, 
Japan Science and Technology, \\
Hongo 5-28-3, Bunkyo-ku, Tokyo 113-0033, Japan}
\date{  }

%
%  abstract
%
\begin{abstract} 

We find that multidimensional determinants "hyperdeterminants", 
related to entanglement measures (the so-called concurrence or 3-tangle for 
the 2 or 3 qubits, respectively), are derived from a duality between entangled 
states and separable states.
By means of the hyperdeterminant and its singularities, the single copy of 
multipartite pure entangled states is classified into an onion structure of 
every closed subset, similar to that by the local rank in the bipartite 
case. 
This reveals how inequivalent multipartite entangled classes are 
partially ordered under local actions.  
In particular, the generic entangled class of the maximal dimension, 
distinguished as the nonzero hyperdeterminant, does not include the maximally 
entangled states in Bell's inequalities in general (e.g., in the 
\(n\!\geq\!4\) qubits), contrary to the widely known bipartite or \(3\)-qubit 
cases.
It suggests that not only are they never locally interconvertible with 
the majority of multipartite entangled states, but they would have no grounds
for the canonical \(n\)-partite entangled states.
Our classification is also useful for the mixed states.
\end{abstract}
%\keywords{multipartite entanglement, hyperdeterminant}

\maketitle

\section{Introduction}
\label{sec:intro} 
Entanglement is the quantum correlation exhibiting nonlocal (nonseparable)
properties.
It is supposed to be never strengthened, on average, by local operations
and classical communication (LOCC).
In particular, entanglement in {\it multi}-parties is of fundamental interest 
in quantum many-body theory \cite{peres95}, and makes quantum information 
processing (QIP), e.g., distillation protocol, more efficient than that 
relying on entanglement only in {\it two}-parties \cite{bennett+96}.
Here, we classify and characterize the multipartite entanglement which has 
yet to be understood, compared with the bipartite one.

%
% bipartite
%
For the single copy of bipartite pure states on 
\({\mathcal H}({\mathbb C}^{k+1})\!\otimes\!{\mathcal H}({\mathbb C}^{k+1})\),
we are interested in whether a state \(|\Psi\rangle\) can convert to another 
state \(|\Phi\rangle\) by LOCC. 
It is convenient to consider the Schmidt decomposition,
\begin{equation}
\label{eq:schmidt}
|\Psi\rangle = \sum_{i_1,i_2=0}^{k}\! a_{i_1,i_2}|i_1\rangle\otimes|i_2\rangle
=\sum_{j=0}^{k} \lambda_j \; |e_j\rangle \otimes |e'_j\rangle,
\end{equation}
where the computational basis \(|i_j\rangle\) is transformed to a local 
biorthonormal basis \(|e_j\rangle, |e'_j\rangle\), and the Schmidt coefficients
\(\lambda_j\) can be taken as \(\lambda_j \geq 0\).
We call the number of nonzero \(\lambda_j \) the (Schmidt) local rank \(r\).
Then the LOCC convertibility is given by a majorization rule between the 
coefficients \(\lambda_j\) of \(|\Psi\rangle\) and those of \(|\Phi\rangle\) 
\cite{nielsen99}.
This suggests that the structure of entangled states consists of partially 
ordered, continuous classes labeled by a set of \(\lambda_j\).
In particular, \(|\Psi\rangle\) and \(|\Phi\rangle\) belong to the same
class under the LOCC classification if and only if all continuous 
\(\lambda_j\) coincide. 

%
%  SLOCC 
%
Suppose we are concerned with a coarse grained classification
by the so-called stochastic LOCC (SLOCC) \cite{bennett+00,dur+00}, where
we identify \(|\Psi\rangle\) and \(|\Phi\rangle\) that are interconvertible 
back and forth with (maybe different) nonvanishing probabilities. 
This is because \(|\Psi\rangle\) and \(|\Phi\rangle\) are supposed to perform 
the same tasks in QIP although their probabilities of success differ.
Later, we find that this SLOCC classification is still fine grained to 
classify the multipartite entanglement. 
Mathematically, two states belong to the same class under SLOCC if and only if
they are converted by an {\it invertible} local operation having a nonzero 
determinant \cite{dur+00}.
Thus the SLOCC classification is equivalent to the classification of orbits 
of the natural action: direct product of general linear groups
\(GL_{k+1}({\mathbb C})\!\times\! GL_{k+1}({\mathbb C})\) \cite{note1}.
The local rank \(r\) in Eq.(\ref{eq:schmidt}) \cite{note2}, equivalently 
the rank of \(a_{i_1,i_2}\), is found to be preserved under SLOCC. 
A set \(S_j\) of states of the local rank \(\leq j\) is a {\it closed} 
subvariety under SLOCC and \(S_{j-1}\) is the singular locus of \(S_j\). 
This is how the local rank leads to an "onion" structure (mathematically 
the stratification):
\begin{equation}
\label{eq:onion}
S_{k+1} \supset S_{k}\supset\cdots\supset S_1 \supset S_0 \!=\! \emptyset, 
\end{equation}
and \(S_{j}\!-\! S_{j-1}\;(j=1,\ldots,k\!+\!1)\) give \(k\!+\!1\) classes of 
entangled states.
Since the local rank can decrease by {\it noninvertible} local operations, 
i.e., general LOCC \cite{lo+01,bennett+00,dur+00}, these classes are totally
ordered such that, in particular, the outermost generic set 
\(S_{k+1}\!-\!S_{k}\) is the class of maximally entangled states and the 
innermost set \(S_{1}(\!=\! S_{1}\!-\!S_{0})\) is that of separable states.

%
%  troubles and motivations in multipartite case
%
For the single copy of multipartite pure states,
\begin{equation}
\label{eq:multi}
|\Psi\rangle = \sum_{i_1,\ldots,i_n =0}^{k_1,\ldots,k_n} 
a_{i_1,\ldots,i_n}|i_1\rangle \otimes \cdots \otimes |i_n\rangle,
\end{equation}
there are difficulties in extending the Schmidt decomposition for 
a multiorthonormal basis \cite{linden+98}.
Moreover, an attempt to use the tensor rank of \(a_{i_1,\ldots,i_n}\) 
\cite{note3} falls down since \(S_j\), defined by it, is not always closed 
\cite{acin+01,brylinski00}.
In the 3 qubits, D\"{u}r {\it et al.} showed that SLOCC classifies 
the whole states \(M\) into {\it finite} classes and in particular there 
exist two inequivalent, Greenberger-Horne-Zeilinger (GHZ) and W, classes of 
the tripartite entanglement \cite{dur+00}. 
They also pointed out that this case is exceptional since
the action \(GL_{k_{1}+1}({\mathbb C})\!\times\!\cdots\!\times\! 
GL_{k_{n}+1}({\mathbb C})\) has {\it infinitely many} orbits in general 
(e.g., for \(n\geq 4\)).

In this paper, we classify multipartite entanglement in a unified manner 
based on the hyperdeterminant. The advantages are three-fold. 
(i) This classification is equivalent to the SLOCC classification when 
SLOCC has finitely many orbits. So it naturally includes the widely known 
bipartite and \(3\)-qubit cases.
(ii) In the multipartite case, we need further SLOCC invariants in addition
to the local ranks. For example, in the \(3\)-qubit case \cite{dur+00}, 
the \(3\)-tangle \(\tau\), just the absolute value of the hyperdeterminant 
(see Eq.~(\ref{eq:tau})), is utilized to distinguish GHZ and W classes.
This work clarifies why the \(3\)-tangle \(\tau\) appears and how these SLOCC
invariants are related to the hyperdeterminant in general.
(iii) Our classification is also useful to multipartite mixed states.
A mixed state \(\rho\) can be decomposed as a convex combination of 
projectors onto pure states.
Considering how \(\rho\) needs at least the outer class in the onion 
structure of pure states, we can also classify multipartite mixed states
into the totally ordered classes (for details, see Appendix~\ref{sec:mixed}). 
We concentrate on the pure states here.

%
%  organization of the paper
%
The rest of the paper is organized as follows.
In Sec.~\ref{sec:2}, a duality between separable states and 
entangled states is introduced. We find that the hyperdeterminant, associated
to this duality, and its singularities lead to the SLOCC-invariant 
onion-like structure of multipartite entanglement.
The characteristics of the hyperdeterminant and its singularities are 
explained in Sec.~\ref{sec:3}.
Classifications of multipartite entangled states are exemplified in 
Sec.~\ref{sec:4} so as to reveal how they are ordered under SLOCC. 
Finally, the conclusion is given in Sec.~\ref{sec:conclude}.

%%%%%%%%%%%%%%%%%%%%%%%%%%%%%%%%%%%%%%%%%%%%%%%%%%%%%%%%%%%%%%%%%%%%%%%%%%%%%%%
\section{Duality between separable states and entangled states}
\label{sec:2}

In this section, we find that there is a duality between the set of 
separable states and that of entangled states.
This duality derives the hyperdeterminant our classification is 
based on.

\subsection{Preliminary: Segre variety}
\label{sec:segre}
%
%
%  geometry  
%
To introduce our idea, we first recall the geometry of pure states.
In a complex (finite) \(k\!+\!1\)-dimensional Hilbert space 
\({\mathcal H}({\mathbb C}^{k+1})\), let \(|\Psi\rangle\) be a (not 
necessarily normalized) vector given by \(k\!+\!1\)-tuple of complex 
amplitudes \(x_{j}(j=0,\ldots,k)\in{\mathbb C}^{k+1}\!-\!\{0\}\) in 
a computational basis.  
The physical state in \({\mathcal H}({\mathbb C}^{k+1})\) is a ray, 
an equivalence class of vectors up to an overall nonzero complex number.
Then the set of rays constitutes the complex projective space 
\({\mathbb C}P^{k}\) (the projectivization of 
\({\mathcal H}({\mathbb C}^{k+1})\)), and \(x := (x_{0}:\ldots : x_{k})\), 
considered up to a complex scalar multiple, gives homogeneous coordinates in 
\({\mathbb C}P^{k}\).

%
%  bipartite,  Segre embedding
%
For a composite system which consists of 
\({\mathcal H}({\mathbb C}^{k_{1}+1})\) and 
\({\mathcal H}({\mathbb C}^{k_{2}+1})\),
the whole Hilbert space is the tensor product 
\({\mathcal H}({\mathbb C}^{k_{1}+1})\otimes
{\mathcal H}({\mathbb C}^{k_{2}+1})\)
and the associated projective space is 
\(M={\mathbb C}P^{(k_{1}+1)(k_{2}+1)-1}\).   
A set \(X\) of the separable states is the mere Cartesian product
\({\mathbb C}P^{k_{1}}\times{\mathbb C}P^{k_{2}}\), whose dimension 
\(k_{1}\!+\!k_{2}\) is much smaller than that of the whole space \(M\),
\((k_{1}\!+\!1)(k_{2}\!+\!1)\!-\!1\).
This \(X\) is a closed, smooth algebraic subvariety (Segre variety) defined by 
the Segre embedding into \({\mathbb C}P^{(k_{1}+1)(k_{2}+1)-1}\) 
\cite{miyake+01,brylinski00},
\begin{gather}
{\mathbb C}P^{k_{1}}\times {\mathbb C}P^{k_{2}} \hookrightarrow
{\mathbb C}P^{(k_{1}+1)(k_{2}+1)-1} , \nonumber \\
\begin{align}
&\left((x^{(1)}_0:\ldots :x^{(1)}_{k_1}),
(x^{(2)}_0:\ldots :x^{(2)}_{k_2})\right)  \\
&\mapsto (x^{(1)}_0 x^{(2)}_0 : \ldots : x^{(1)}_0 x^{(2)}_{k_2} : 
x^{(1)}_1 x^{(2)}_0 : \ldots\ldots : x^{(1)}_{k_1} x^{(2)}_{k_2}). \nonumber
\end{align}
\end{gather}
Denoting homogeneous coordinates in \({\mathbb C}P^{(k_{1}+1)(k_{2}+1)-1}\)
by \(b_{i_1,i_2}\!=\! x^{(1)}_{i_1}x^{(2)}_{i_2} \; (0\leq i_j \leq k_j)\), 
we find that the Segre variety \(X\) is given by the common zero locus of 
\(k_{1}(k_{1}\!+\!1)k_{2}(k_{2}\!+\!1)/4\) 
homogeneous polynomials of degree \(2\):
\begin{equation}
b_{i_{1},i_{2}} b_{i'_{1}, i'_{2}} - b_{i_{1},i'_{2}} b_{i'_{1}, i_{2}},
\end{equation}
where \(0 \leq i_{1} < i'_{1} \leq k_1, \; 0 \leq i_{2} < i'_{2} \leq k_2\).
Note that this condition implies that all \(2\times 2\) minors of "matrix" 
\(b_{i_1,i_2}\) equal 0; i.e., the rank of \(b_{i_1,i_2}\) is 1.
Thus we have \(X=S_1\), which agrees with the SLOCC classification by 
the local rank in the bipartite case.

%
%  multipartite
%
Now consider the multipartite Cartesian product 
\(X\!=\!{\mathbb C}P^{k_{1}}\!\times\!\cdots\!\times\!{\mathbb C}P^{k_{n}}\) 
in the Segre embedding into 
\(M\!=\!{\mathbb C}P^{(k_{1}+1)\cdots(k_{n}+1)-1}\).
Because this Segre variety \(X\) is the projectivization of a variety of 
the matrices \(b_{i_1,\ldots, i_n}\!=\! x^{(1)}_{i_1}\cdots x^{(n)}_{i_n}\), 
it gives a set of the completely separable states
in \({\mathcal H}({\mathbb C}^{k_{1}+1})\!\otimes\!\cdots\!\otimes\! 
{\mathcal H}({\mathbb C}^{k_{n}+1})\). 
By another Segre embedding, say \(X'\!=\!{\mathbb C}P^{(k_{1}+1)(k_{2}+1)-1}
\!\times\!{\mathbb C}P^{k_3}\!\times\!\cdots\!\times\!{\mathbb C}P^{k_n}\), 
we also distinguish a set of separable states where only 1st and 2nd 
parties can be entangled, i.e., when we regard 1st and 2nd parties as 
one party, an element of this set is completely separable for 
"\(n\!-\!1\)" parties.
This is how, also in the multipartite case, we can classify all kinds of 
{\it separable} states, typically lower dimensional sets.
Note that, in the multipartite case, this check for the separability is 
more strict than the check by local ranks \cite{note4}.

\subsection{Main idea: duality}
\label{sec:dual}
We rather want to classify {\it entangled} states, typically higher 
dimensional complementary sets of separable states.
Our strategy is based on the duality in algebraic geometry; a hyperplane in 
\({\mathbb C}P\) forms the point of a dual projective space 
\({\mathbb C}P^{\ast}\), and conversely every point \(p\) of \({\mathbb C}P\) 
is tied to a hyperplane \(p^{\vee}\) in \({\mathbb C}P^{\ast}\) as 
the set of all hyperplanes in \({\mathbb C}P\) passing through \(p\).
Remarkably, the projective duality between projective subspaces, like the 
above example, can be extended to an involutive correspondence between 
irreducible algebraic subvarieties in \({\mathbb C}P\) and 
\({\mathbb C}P^{\ast}\).
So we define a projectively dual (irreducible) variety \(X^{\vee} \subset 
{\mathbb C}P^{\ast}\) as the closure of the set of all hyperplanes tangent 
to the Segre variety \(X\).

Let us observe (and see the reason later) that, in the bipartite case seen in 
Sec.~\ref{sec:intro}, the variety \(S_{k}\) of the {\it degenerate} 
\((k\!+\!1) \times (k\!+\!1)\) matrices \(A \!=\! a_{i_1,i_2}\) is 
projectively dual to the variety \(S_1 \!=\! X\) of the matrices 
\(B \!=\! b_{i_1,i_2} \!=\! x^{(1)}_{i_1} x^{(2)}_{i_2}\).
That is, \(S_{k}\) is the dual variety \(X^{\vee}\). 
Following an analogy with a 2-dimensional (bipartite) case, 
an \(n\)-dimensional matrix \(A \!=\! a_{i_1,\ldots,i_n}\) is called 
{\it degenerate} if and only if it (precisely, its projectivization) lies 
in the projectively dual variety \(X^{\vee}\) of the Segre variety \(X\).
In other words, identifying the space of \(n\)-dimensional matrices with 
its dual by means of the pairing,
\begin{equation}
\label{eq:F}
F(A,B)
\!=\!\!\!\sum_{i_1,\ldots, i_n =0}^{k_1,\ldots,k_n}
   \!\! a_{i_1,\ldots, i_n} b_{i_1,\ldots, i_n},
\end{equation}
we see that \(A\) is degenerate if and only if  
its orthogonal hyperplane \(F(A,B) \!=\! 0\) is tangent to \(X\) at some 
nonzero point \(x=(x^{(1)},\ldots, x^{(n)})\). Analytically, a set of 
equations, 
\begin{equation}
\label{eq:critical}
\left\{\mbox{  }
\begin{split}
& F(A,x) = \sum_{i_1,\ldots, i_n =0}^{k_1,\ldots,k_n} 
a_{i_1,\ldots, i_n} x^{(1)}_{i_1}\cdots x^{(n)}_{i_n} =  0 , \\
& \frac{\partial}{\partial x^{(j)}_{i_j}} F(A,x) = 0 
\quad \mbox{for all} \; j,i_j  
\end{split}
\right. 
\end{equation}
(\(j=1,\ldots,n\) and \(0 \leq i_j \leq k_j\)),
has at least a nontrivial solution \(x \!=\! (x^{(1)},\ldots,x^{(n)})\) of 
every \(x^{(j)} \!\ne\! 0\), and then \(x\) is called a critical point.
The above condition is also equivalent to saying that 
the kernel \({\rm ker} F\) of \(F(A,x)\) is not empty,
where \({\rm ker} F\) is the set of points 
\(x=(x^{(1)},\ldots,x^{(n)}) \in X\) such that, in every \(j_0 = 1,\ldots, n\),
\begin{equation}
F \mbox{\mathversion{bold}$($}
A,(x^{(1)},\ldots,x^{(j_0 -1)},z^{(j_0)},x^{(j_0 +1)},\ldots,x^{(n)})
\mbox{\mathversion{bold}$)$}=0,
\end{equation} 
for the {\it arbitrary} \(z^{(j_0)}\).

In the case of \(n\!=\!2\), the condition for Eqs.(\ref{eq:critical}) 
coincides with the usual notion of degeneracy and 
means that \(A\) does not have the full rank.
It shows that \(X^{\vee}\) is nothing but \(S_{k}\).  
In particular, \(X^{\vee}\), defined by this condition, is of codimension 1 
and is given by the ordinary determinant \(\det A \!=\! 0\), if and only if 
\(A\) is a square \((k_{1} \!=\! k_{2} \!=\! k)\) matrix.
In the \(n\)-dimensional case, if \(X^{\vee}\) is a hypersurface 
(of codimension 1), it is given by the zero locus of a unique (up to sign) 
irreducible homogeneous polynomial over \({\mathbb Z}\) of 
\(a_{i_{1},\ldots,i_{n}}\).
This polynomial is the hyperdeterminant introduced by Cayley and 
is denoted by \({\rm Det} A\).
As usual, if \(X^{\vee}\) is not a hypersurface, we set \({\rm Det} A\)
to be 1. 

Remember that, in the {\it bipartite} case, 
we classify the states \(\in S_{k+1}\!-\!S_{k} \!=\!  M\!-\!X^{\vee}\) as 
the generic entangled states, the states 
\(\in S_{k}\!-\!S_{k-1} \!=\!  X^{\vee}\!-\!X^{\vee}_{\rm sing}\) as the 
next generic entangled states, and so on.
Likewise, we aim to classify the {\it multipartite} entangled states into the 
onion structure by the dual variety \(X^{\vee}\) 
(\({\rm Det}A\!=\!0\)), its singular locus \(X^{\vee}_{\rm sing}\) and so on
(i.e., by every closed subvariety), instead of the tensor rank \cite{note3}.

%%%%%%%%%%%%%%%%%%%%%%%%%%%%%%%%%%%%%%%%%%%%%%%%%%%%%%%%%%%%%%%%%%%%%%%%%%%%%%%
\section{Hyperdeterminant and its singularities }
\label{sec:3}
In order to classify multipartite entanglement into the SLOCC-invariant onion 
structure, we explore the dual variety \(X^{\vee}\) (zero hyperdeterminant) 
and its singular locus in this section.

\subsection{Hyperdeterminant} 
\label{sec:hdet}
We utilize the hyperdeterminant, the generalized determinant for 
higher dimensional matrices by Gelfand {\it et al.}
\cite{gelfand+92,gelfand+94}.
Its absolute value is also known as an entanglement measure,
the concurrence \(C\) \cite{hill+97} or 3-tangle \(\tau\) \cite{coffman+00} 
for the 2 or 3-qubit pure case, respectively.
\begin{align}
\label{eq:C}
C &= 2|{\rm Det} A_2| = 2|\det A|= 2|a_{00}a_{11}-a_{01}a_{10}|, \\
\label{eq:tau}
\tau &= 4|{\rm Det}A_3| \nonumber\\
     &= 4| a_{000}^2 a_{111}^2 + a_{001}^2 a_{110}^2 +
a_{010}^2 a_{101}^2 + a_{100}^2 a_{011}^2 \nonumber\\ 
&\quad -2(a_{000}a_{001}a_{110}a_{111}+a_{000}a_{010}a_{101}a_{111}\nonumber\\ 
&\quad + a_{000}a_{100}a_{011}a_{111}+a_{001}a_{010}a_{101}a_{110} \nonumber\\
&\quad + a_{001}a_{100}a_{011}a_{110}+a_{010}a_{100}a_{011}a_{101}) \nonumber\\
&\quad + 4 (a_{000}a_{011}a_{101}a_{110} + a_{001}a_{010}a_{100}a_{111})|.
\end{align}

%
%  characteristics
%
The following useful facts are found in \cite{gelfand+94}.
Without loss of generality, we assume that \(k_{1}\!\geq\! k_{2} \!\geq\!
\cdots\!\geq\! k_{n} \!\geq\! 1\).
The \(n\)-dimensional hyperdeterminant \({\rm Det} A\) of format 
\((k_{1}\!+\!1)\!\times\!\cdots\!\times\!(k_{n}\!+\!1)\) exists, 
i.e., \(X^{\vee}\) is a hypersurface, if and only if a "polygon inequality"
\(k_1 \!\leq\! k_2 + \cdots\!+\! k_n\) is satisfied. 
For \(n\!=\!2\), this condition is reduced to \(k_{1} \!=\! k_{2}\) 
as desired, and \({\rm Det} A\) coincides with \(\det A\). 
The matrix format is called boundary if \(k_{1} \!=\! k_{2} \!+ \cdots 
+\! k_{n}\) and interior if \(k_{1} \!<\! k_{2} \!+\cdots+\! k_{n}\).
Note that (i) The boundary format includes the "bipartite cut" between 
the 1st party and the others so that it is mathematically tractable. 
(ii) the interior format includes the \(n \!\geq\! 3\)-qubit case.
We treat hereafter the format where the polygon inequality holds and  
\(X^{\vee}\) is the largest closed subvariety, defined by the hypersurface 
\({\rm Det}A =0\).

\({\rm Det} A\) is relatively invariant (invariant up to constant) under the 
action of 
\(GL_{k_1 + 1}({\mathbb C})\times\cdots\times GL_{k_n + 1}({\mathbb C})\).
In particular, interchanging two parallel slices (submatrices with some fixed 
directions) leaves \({\rm Det} A\) invariant up to sign, and \({\rm Det} A\) 
is a homogeneous polynomial in the entries of each slice. 
Since it is ensured that \(X^{\vee}\), \(X^{\vee}_{\rm sing}\), and further
singularities are invariant under SLOCC, our classification is equivalent to 
or coarser than the SLOCC classification. Later, we see that the former and 
the latter correspond to the case where SLOCC gives finitely and infinitely 
many classes, respectively.

\subsection{Schl\"{a}fli's construction}
\label{sec:schlafli}
It would not be easy to calculate \({\rm Det}A\) directly by its definition 
that Eqs.(\ref{eq:critical}) have at least one solution.
Still, the Schl\"{a}fli's method enables us to construct \({\rm Det}A_n\) 
of format \(2^n\) (\(n\) qubits) by induction on \(n\) 
\cite{gelfand+92,gelfand+94,schlafli}.

For \(n\!=\!2\), by definition \({\rm Det}A_{2} \!=\!\det A \!=\!
a_{00}a_{11}\!-\!a_{01}a_{10}\).
Suppose \({\rm Det} A_{n}\), whose degree of homogeneity is \(l\), is given.
Associating an \(n\!+\!1\)-dimensional matrix \(a_{i_0,i_1,\ldots,i_n}\)  
(\(i_{j}\!=\!0,1\)) to a family of \(n\)-dimensional matrices 
\(\widetilde{A}(x) \!=\! \sum_{i_0}a_{i_{0},i_{1},\cdots ,i_{n}}x_{i_0}\) 
linearly depending on the auxiliary variable \(x_{i_0}\),
we have \({\rm Det} \widetilde{A}(x)_{n}\).
Due to Theorem 4.1 and 4.2 of \cite{gelfand+94}, the discriminant \(\Delta\) 
of \({\rm Det} \widetilde{A}(x)_{n}\) gives \({\rm Det} A_{n+1}\) with an extra
factor \(R_n\).
The Sylvester formula of the discriminant \(\Delta\) for binary forms enables 
us to write \({\rm Det}A_{n+1}\) in terms of the determinant of order 
\(2l\!-\!1\);
\begin{align}
\label{eq:schlafli}
&{\rm Det}A_{n+1}
=\Delta ({\rm Det} \widetilde{A}(x)_{n})/R_{n} \nonumber\\ 
&=\frac{1}{R_n c_l} \left| \begin{array}{cccccccc}
c_0    & c_1   &\cdots& c_{l-2}  & c_{l-1} & c_l    &\cdots& 0\\
0      & c_0   &\cdots& \cdots   & c_{l-2} & c_{l-1}&\cdots& 0\\
\vdots &       &\ddots&          & \vdots  &        &\ddots& \vdots\\
0      & 0     &\cdots&  c_0     & c_1     & \cdots &\cdots& c_l \\
1 c_1  & 2 c_2 &\cdots& \cdots   & l c_l   & 0      &\cdots& 0\\
\vdots &       &\ddots&          & \vdots  &        &\ddots& \vdots\\
0      & 0     &\cdots& 0        & 1 c_1   & 2 c_2  &\cdots& l c_l\\ 
\end{array}\right|,
\end{align}
where each \(c_j\) is the coefficient of \(x_{0}^{l-j} x_{1}^j\) in 
\({\rm Det} \widetilde{A}(x)_{n}\), i.e., 
\(c_j = \frac{1}{(l-j)! j!} 
\frac{\partial^{l}}{\partial x_{0}^{l-j} \partial x_{1}^{j}} 
 {\rm Det} \widetilde{A}(x)_{n}\). 

Note that because for \(n\!=\! 2\) or \(3\), the extra factor \(R_n\) is just 
a nonzero constant, \({\rm Det}A_{3,4}\) for the 3 or 4 qubits is readily 
calculated, respectively.
It would be instructive to check that \({\rm Det}A_3\) in Eq.(\ref{eq:tau}) 
is obtained in this way.
On the other hand, for \(n\!\geq\! 4\), \(R_n\) is the Chow form (related 
resultant) of irreducible components of the singular locus 
\(X^{\vee}_{\rm sing}\).
These are due to the fact that \(X^{\vee}_{\rm sing}\) has codimension \(2\) 
in \(M\) for any format of the dimension \(n \!\geq\! 3\) except for 
the format \(2^3\) (\(3\)-qubit case), which was conjectured in 
\cite{gelfand+92} and was proved in \cite{weyman+96}.
So we have to explore \(X^{\vee}_{\rm sing}\) not only to classify entangled 
states in the \(n\) qubits, but to calculate \({\rm Det}A_{n+1}\) inductively.
Although \({\rm Det}A_{n\geq 5}\) has yet to be written explicitly, 
only its degree \(l\) of homogeneity is known 
(in Corollary 2.10 of \cite{gelfand+94}) to grow very fast as 
\(2,4,24,128,880,6816,60032,589312,6384384\) for \(n=2,3,\ldots,10\).

\subsection{Singularities of the hyperdeterminant }
\label{sec:sing}
We describe the singular locus of the dual variety \(X^{\vee}\).
The technical details are given in \cite{weyman+96}.
It is known that, for the boundary format, the next largest closed subvariety 
\(X^{\vee}_{\rm sing}\) is always an irreducible hypersurface in \(X^{\vee}\);
in contrast, for the interior one, \(X^{\vee}_{\rm sing}\) has generally 
two closed irreducible components of codimension 1 in \(X^{\vee}\), 
{\it node}-type \(X^{\vee}_{\rm node}\) and {\it cusp}-type 
\(X^{\vee}_{\rm cusp}\) singularities.
The rest of this subsection can be skipped for the first reading.
It is also illustrated for the \(3\)-qubit case in 
Appendix~\ref{sec:rep_3qubit}.

\begin{figure}[t]
\begin{center}
\begin{psfrags}
\psfrag{X}{\large{$X$}}
\psfrag{V}{\large{$X^{\vee}$}}
\includegraphics[clip]{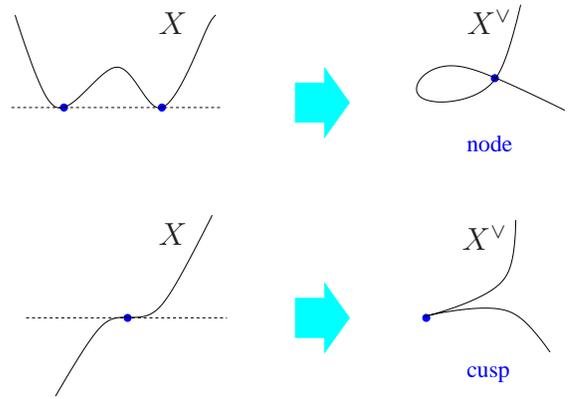}
\caption{
Two types of singularities of \(X^{\vee}\). 
\(X^{\vee}_{\rm node}\) corresponds to the bitangent of \(X\), where
both tangencies are of the first order.
\(X^{\vee}_{\rm cusp}\) corresponds to the tangent at an inflection point 
of \(X\), where its tangency is of the second order. } 
\label{fig:sing_xv}
\end{psfrags}
\end{center}
\end{figure}

First, \(X^{\vee}_{\rm node}\) is the closure of the set of 
hyperplanes tangent to the Segre variety \(X\) at more than one point
(cf. Fig.~\ref{fig:sing_xv}).
\(X^{\vee}_{\rm node}\) can be composed of closed irreducible subvarieties 
\(X^{\vee}_{\rm node}(J)\) labeled by the subset 
\(J \!\subset\! \{1,\ldots,n\},\) including \(\emptyset\). 
Indicating that two solutions \(x\!=\!(x^{(1)},\ldots,x^{(j)},\ldots,
x^{(n)})\) of Eq.(\ref{eq:critical}) coincide for \(j\!\in\! J\), 
the label \(J\) distinguishes the pattern in these solutions. 
In order to rewrite \(X^{\vee}_{\rm node}(J)\), let us pick up a point 
\(x^o(J)\) such that its homogeneous coordinates 
\(x^{(j)}_{i_j} = \delta_{i_j,0}\) for \(j\!\in\! J\) and 
\(\delta_{i_j,k_j}\) for \(j \!\notin\! J\). 
It is convenient to label the positions of \(1\) in each \(x^{(j)}\) 
by a multi-index \([i_1,\ldots,i_n]\).
For example, \(x^o(1)\) is labeled by \([0,k_2,\ldots,k_n]\) and 
\(x^o(1,\ldots,n)\) is just written by \(x^o\).
According to Eqs.(\ref{eq:critical}), \(X^{\vee}|_{x^o(J)}\), tangent to 
\(X\) at \(x^o(J)\), consists of the matrices \(A\) of all  
\(a_{i'_1,\ldots,i'_n} \!=\! 0\) such that \([i'_1,\ldots,i'_n]\) 
differs from \([i_1,\ldots,i_n]\) of \(x^o(J)\) in at most one index.
Then we can define \(X^{\vee}_{\rm node}(J)\) as  
\begin{equation}
\label{eq:node}
X^{\vee}_{\rm node}(J) =
\overline{(X^{\vee}|_{x^o} \cap X^{\vee}|_{x^o(J)})\cdot G},
\end{equation} 
where \(G\!=\!GL_{k_1+1}\!\times\!\cdots\!\times\! GL_{k_n+1}\) acts on \(M\) 
from the right and the bar stands for the closure.

Second, \(X^{\vee}_{\rm cusp}\) is the set of hyperplanes having a critical 
point which is not a simple quadratic singularity (cf. Fig.~\ref{fig:sing_xv}).
Precisely, the quadric part of \(F(A,x)\) at \(x^o\) is a matrix
\(y_{(j,i_j),(j',i_{j'})}\!=\!
(\partial^2 /\partial x_{i_{j}}^{(j)}\partial x_{i_{j'}}^{(j')})F(A,x^o)\), 
where the pairs \((j,i_j),\:(j',i_{j'})\:(1 \!\leq\! i_j \!\leq\! k_j,\: 
1 \!\leq\! i_{j'} \!\leq\! k_{j'})\) are the row and the column index, 
respectively.
Denoting by \(X^{\vee}_{\rm cusp}|_{x^o}\) the variety of the Hessian 
\(\det y \!=\!0\) in \(X^{\vee}|_{x^o}\), we can define 
\(X^{\vee}_{\rm cusp}\) as 
\begin{equation}
\label{eq:cusp}
X^{\vee}_{\rm cusp} = X^{\vee}_{\rm cusp}|_{x^o}\cdot G.
\end{equation}
This \(X^{\vee}_{\rm cusp}\) is already closed without taking the closure.

%%%%%%%%%%%%%%%%%%%%%%%%%%%%%%%%%%%%%%%%%%%%%%%%%%%%%%%%%%%%%%%%%%%%%%%%%%%%%%%
\section{Classification of multipartite entanglement}
\label{sec:4}
According to Sec.~\ref{sec:2} and \ref{sec:3}, we illustrate the classification
of multipartite pure entangled states for typical cases.

\subsection{3-qubit (format \(2^3\)) case}
\label{sec:3qubit}
\begin{figure}[t]
\begin{center}
\begin{psfrags}
\psfrag{m}{$M$}
\psfrag{v}{$X^{\vee}$}
\psfrag{s}{$X^{\vee}_{\rm cusp}$}
\psfrag{a}{\color{red}$X^{\vee}_{\rm node}(1)$ }
\psfrag{b}{\color{blue}$X^{\vee}_{\rm node}(2)$}
\psfrag{c}{\color{dgreen}$X^{\vee}_{\rm node}(3)$}
\psfrag{x}{$X$}
\psfrag{G}{\rotatebox{60}{GHZ}}
\psfrag{W}{\rotatebox{60}{W}}
\includegraphics[clip]{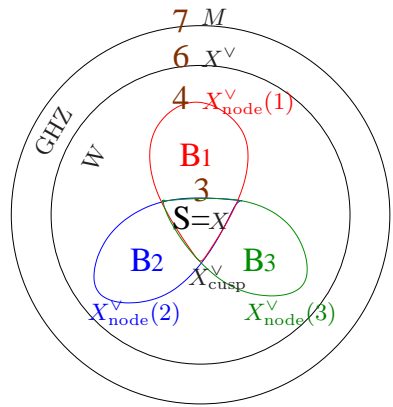}
\caption{
The onion-like classification of SLOCC orbits in the \(3\)-qubit case.
We utilize a duality between the smallest closed subvariety 
\(X\) and the largest closed subvariety \(X^{\vee}\).
The dual variety \(X^{\vee}\) (zero hyperdeterminant) and 
its singularities constitute SLOCC-invariant closed subvarieties, 
so that they classify the multipartite entangled states (SLOCC orbits).}
\label{fig:onion_3bit}
\end{psfrags}
\end{center}
\end{figure}

The classification of the 3 qubits under SLOCC has been already done in 
\cite{dur+00,note1}.
Surprisingly, Gelfand {\it et al.} considered the same mathematical problem 
by \({\rm Det}A_3\) in Example 4.5 of \cite{gelfand+94}. 
Our idea is inspired by this example.
We complement the Gelfand {\it et al.}'s result, analyzing additionally  
the singularities of \(X^{\vee}\) in detail. 
The dimensions, representatives, names, and varieties of the orbits are 
summarized as follows. 
The basis vector \(|i_1\rangle\otimes | i_2 \rangle\otimes |i_3\rangle\) 
is abbreviated to \(|i_1 i_2 i_3 \rangle\).
\bigskip \\
dim 7:  \(|000\rangle + |111\rangle,\)       GHZ  
        \(\in\: M(={\mathbb C}P^7)\!-\!X^{\vee}\). \\
dim 6:  \(|001\rangle + |010\rangle + |100\rangle,\) W  
        \(\in\: X^{\vee}\!-\!X^{\vee}_{\rm sing}
        =X^{\vee}\!-\!X^{\vee}_{\rm cusp}\). \\
dim 4:  \(|001\rangle + |010\rangle,\: |001\rangle + |100\rangle,\: 
        |010\rangle + |100\rangle,\) \\ 
\quad   biseparable \(B_{j}\) \(\in\: X^{\vee}_{\rm node}(j) \!-\! X\) 
        for \(j=1,2,3\). \\
\quad   \(X^{\vee}_{\rm node}(j)={\mathbb C}P^1_{j{\rm \mbox{-}th}} \!\times\! 
        {\mathbb C}P^3\) are three closed irreducible \\ 
\quad   components of \(X^{\vee}_{\rm sing}=X^{\vee}_{\rm cusp}\).\\
dim 3:  \(|000\rangle,\) completely separable \(S\) \\
\quad   \(\in\: X= \bigcap_{j=1,2,3} X^{\vee}_{\rm node}(j) 
        ={\mathbb C}P^1 \!\times\!{\mathbb C}P^1 \!\times\!{\mathbb C}P^1\). 
\bigskip  

\(G\!=\!GL_2 \times GL_2 \times GL_2\) has the onion structure of six orbits 
on \(M\) (see Fig.~\ref{fig:onion_3bit}), by excluding the orbit 
\(\emptyset (\!=\!X^{\vee}_{\rm node}(\emptyset))\).
The dual variety \(X^{\vee}\) is given by \({\rm Det}A_3 =0\) 
(cf. Eq.~(\ref{eq:tau})). 
Its dimension is \(7-1=6\).
The outside of \(X^{\vee}\) is generic tripartite entangled class of 
the maximal dimension, whose representative is GHZ.
This suggests that almost any state in the \(3\) qubits can be locally 
transformed into GHZ with a finite probability, and vice versa.
Next, we can identify \(X^{\vee}_{\rm sing}\) as \(X^{\vee}_{\rm cusp}\), 
which is the union of three closed irreducible subvarieties 
\(X^{\vee}_{\rm node}(j)\) for \(j\!=\!1,2,3\) \cite{weyman+96} (also see 
Appendix~\ref{sec:rep_3qubit}). 
For example, \(X^{\vee}_{\rm node}(1)\) means by definition that, 
in addition to the condition of \(X^{\vee}\) in Sec.~\ref{sec:dual},   
there exists some nonzero \(x^{(1)}\) such that \(F(A,x)\!=\!0\)  
for any \(x^{(2)},x^{(3)}\); i.e., a set of linear equations 
\(y_{i_2,i_3}(x^{(1)})\!=\!
(\partial^2/\partial x^{(2)}_{i_2}\partial x^{(3)}_{i_3})F(A,x)\!=\!0\) for
\(i_j\!=\!0,1\) has a nontrivial solution \(x^{(1)}\).
This indicates that the "bipartite" matrix
\begin{equation}
\label{eq:B1}
\left(\begin{array}{cccc}
a_{000} & a_{001} & a_{010} & a_{011} \\
a_{100} & a_{101} & a_{110} & a_{111}
\end{array}\right)
\end{equation}
never has the full rank  (i.e., six \(2 \!\times\! 2\) minors in 
Eq.(\ref{eq:B1}) are zero). We can identify \(X^{\vee}_{\rm node}(1)\) as 
the set \({\mathbb C}P^{1}_{1{\rm st}}\times{\mathbb C}P^{3}\), seen in 
Sec.~\ref{sec:segre}, of biseparable states between the 1st party and 
the rest of the parties.
Its dimension is \(1+3=4\).
Likewise, \(X^{\vee}_{\rm node}(j)\) for \(j=2,3\) gives the biseparable
class for the 2nd or 3rd party, respectively.
So, the class of \(X^{\vee}\!-\! X^{\vee}_{\rm sing}\) is found to be 
tripartite entangled states, whose representative is W.
We can intuitively see that, among genuine tripartite entangled states, 
W is rare, compared to GHZ \cite{dur+00}.
Finally, the intersection of \(X^{\vee}_{\rm node}(j)\) is the completely 
separable class \(S\), given by the Segre variety \(X\) of dimension \(3\).
Another intuitive explanation about this procedure is seen in 
Appendix~\ref{sec:rep_3qubit}.

Now we clarify the relationship of six classes by {\it noninvertible} local
operations.
Because noninvertible local operations cause the decrease in local ranks 
\cite{note5}, the partially ordered structure of entangled states in the 
\(3\) qubits, included in Fig.~\ref{fig:part_order}, appears.
Two inequivalent tripartite entangled classes, GHZ and W, have the same local 
ranks \((2,2,2)\) for each party so that they are not interconvertible 
by the noninvertible local operations (i.e., general LOCC).
Two classes hold different physical properties \cite{dur+00}; the GHZ 
representative state has the maximal amount of generic tripartite entanglement
measured by the \(3\)-tangle \(\tau \!\propto\! |{\rm Det}A_3|\), while the W 
representative state has the maximal amount of (average) \(2\)-partite 
entanglement distributed over \(3\) parties (also \cite{koashi+00}). 
Under LOCC, a state in these two classes can be transformed into any state 
in one of the three biseparable classes \(B_j \;(j=1,2,3)\), where
the \(j\)-th local rank is \(1\) and the others are \(2\). 
Three classes \(B_j\) never convert into each other.
Likewise, a state in \(B_j\) can be locally transformed into any state 
in the completely separable class \(S\) of local ranks \((1,1,1)\).

This is how the onion-like classification of SLOCC orbits reveals that 
multipartite entangled classes constitute the partially ordered structure.
It indicates significant differences from the totally ordered one in the 
bipartite case.
(i) In the \(3\)-qubit case, all SLOCC invariants we need to classify is 
the hyperdeterminant \({\rm Det}A_3\) in addition to local ranks.
(ii) Although noninvertible local operations generally mean
the transformation further inside the onion structure, 
an outer class can not necessarily be transformed into the {\it neighboring} 
inner class. A good example is given by GHZ and W, as we have just seen.

%%%%%%%%%%%%%%%%%%%%%%%%%%%%%%%%%%%%%%%%%%%%%%%%%%%%%%%%%%%%%%%%%%%%%%%%%%%%%%%
%
\subsection{Format \({\rm 3\times 2\times 2} \) case}
\label{sec:3x2x2}
Before proceeding to the \(n\!\geq\! 4\)-qubit case, we drop in the format 
\(3 \!\times\! 2 \!\times\! 2\), which would give an insight into 
the structure of multipartite entangled states when each party has 
a system consisting of more than two levels. 
This case is interesting since on the one hand (contrary to the \(3\)-qubit 
case), it is typical that GHZ and W are included in \(X^{\vee}_{\rm sing}\); 
on the other hand (similarly to the bipartite or \(3\)-qubit cases),
SLOCC has still finite classes so that it becomes another good test for the 
equivalence to the SLOCC classification.
Besides, it is a boundary format so that several subvarieties can be 
explicitly calculated, and enables us to analyze entanglement in the qubits 
system using an auxiliary level, like ion traps.  
\bigskip \\
dim 11: \(|000\rangle + |101\rangle + |110\rangle + |211\rangle\) 
        \(\in\: M(={\mathbb C}P^{11})\!-\!X^{\vee}\). \\
dim 10: \(|000\rangle + |101\rangle + |211\rangle\)
        \(\in\: X^{\vee}\!-\!X^{\vee}_{\rm sing}
        =X^{\vee}\!-\!X^{\vee}_{\rm node}(1)\).\\  
dim 9:  \(|000\rangle + |111\rangle\), GHZ 
        \(\in\: X^{\vee}_{\rm sing}(\!=\!X^{\vee}_{\rm node}(1))
                \!-\!X^{\vee}_{\rm cusp}\). \\
dim 8:  \(|001\rangle + |010\rangle + |100\rangle\), W   
        \(\in\: X^{\vee}_{\rm cusp} \!-\! 
          \bigcup_{j=\emptyset,2,3}X^{\vee}_{\rm node}(j) \).\\
dim 6:  \(|001\rangle + |100\rangle,\: |010\rangle + |100\rangle,\) 
        biseparable \(B_2,\: B_3\)\\ 
\quad   \(\in\: X^{\vee}_{\rm node}(2) \!-\! X,\:
                X^{\vee}_{\rm node}(3) \!-\! X\). \\
dim 5:  \(|001\rangle + |010\rangle\), biseparable \(B_1\) 
        \(\in\: X^{\vee}_{\rm node}(\emptyset) \!-\! X\). \\
dim 4:  \(|000\rangle\), completely separable \(S\) \\ 
\quad   \(\in X = {\mathbb C}P^{2}\!\times{\mathbb C}P^{1}
        \!\times{\mathbb C}P^{1}\). 
\bigskip

\begin{figure}[t]
\begin{center}
\begin{psfrags}
\psfrag{m}{$M$}
\psfrag{v}{$X^{\vee}$}
\psfrag{g}{$X^{\vee}_{\rm node}(1)$}
\psfrag{w}{$X^{\vee}_{\rm cusp}$}
\psfrag{a}{{\color{red}$X^{\vee}_{\rm node}(\emptyset)$}}
\psfrag{b}{{\color{blue}$X^{\vee}_{\rm node}(2)$}}
\psfrag{c}{{\color{dgreen}$X^{\vee}_{\rm node}(3)$}}
\psfrag{x}{$X$}
\psfrag{F}{\rotatebox{60}
{$|000\rangle\!+\!|101\rangle\!+\!|110\rangle\!+\!|211\rangle$}}
\psfrag{T}{\rotatebox{60}{$|000\rangle\!+\!|101\rangle\!+\!|211\rangle$}}
\psfrag{G}{\rotatebox{60}{GHZ}}
\psfrag{W}{\rotatebox{60}{W}}
\includegraphics[clip]{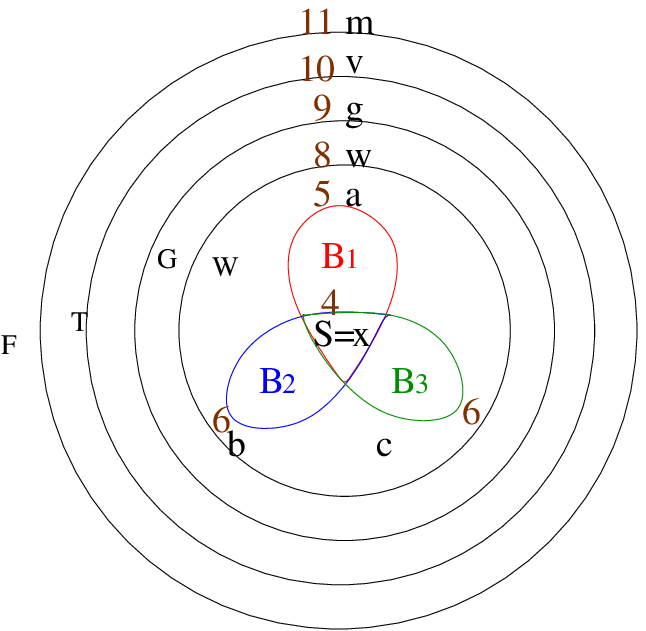}
\caption{ 
The onion-like classification of SLOCC orbits in the 
\(3 \!\times\! 2 \!\times\! 2\) format.
Although this resembles Fig.~\ref{fig:onion_3bit} in the order of SLOCC 
orbits (two orbits are added outside), it is worth while to note that 
singularities of \(X^{\vee}\), which classify the SLOCC orbits, have 
a different order. } 
\label{fig:onion_3x2x2}
\end{psfrags}
\end{center}
\begin{center}
\includegraphics[clip]{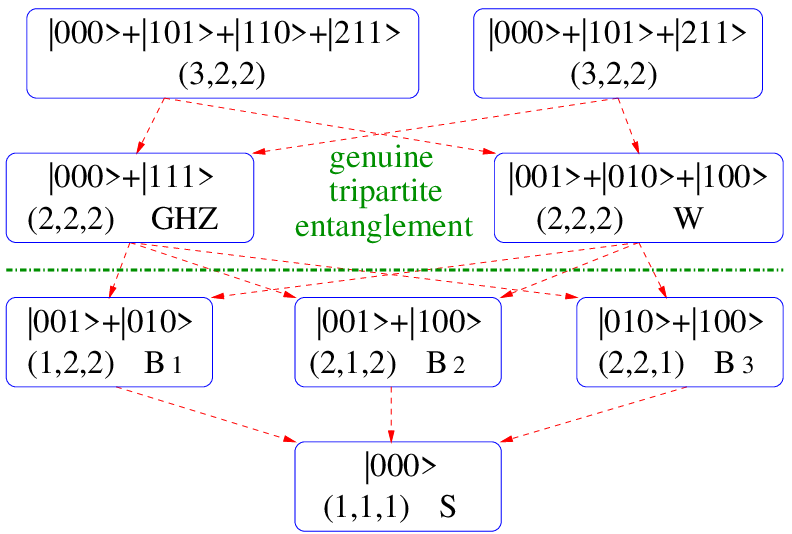}
\caption{
The partially ordered structure of multipartite pure entangled states
in the \(3\times 2\times 2\) format, including the \(3\)-qubit case.
Each class, corresponding to the SLOCC orbit, is labeled by 
the representative, local ranks, and the name.
Noninvertible local operations, indicated by dashed arrows, 
degrade "higher" entangled classes into "lower" entangled ones.}  
\label{fig:part_order}
\end{center}
\end{figure}

The onion structure consists of eight orbits on \(M\) under SLOCC 
(see Fig.~\ref{fig:onion_3x2x2}).
Generic entangled states of the outermost class are given by nonzero 
\({\rm Det}A\), which can be calculated in the {\it boundary} format as the 
determinant associated with the Cayley-Koszul complex. 
Although this is one of the Gelfand {\it et al.}'s recent successes for 
generalized discriminants, we avoid its detailed explanation here.
According to Theorem 3.3 of \cite{gelfand+94}, we have 
\begin{equation}
\label{eq:DetA_3x2x2}
{\rm Det}A = m_1 m_4 -m_2 m_3
\end{equation}
of degree \(6\), where \(m_j \;(j=1,2,3,4)\) is the \(3 \times 3\) minor of 
\begin{equation}
\label{eq:A_3x2x2}
\left( \begin{array}{cccc}
a_{000} & a_{001} & a_{010} & a_{011} \\
a_{100} & a_{101} & a_{110} & a_{111} \\
a_{200} & a_{201} & a_{210} & a_{211}
       \end{array} \right) 
\end{equation}
without the \(j\)-th column, respectively.
Next, it is characteristic that \(X^{\vee}_{\rm sing}\) is 
\(X^{\vee}_{\rm node}(1)\) \cite{weyman+96}. 
Similarly to the \(3\)-qubit case in Sec.~\ref{sec:3qubit}, 
\(X^{\vee}_{\rm node}(1)\) means that the "bipartite" matrix in 
Eq.~(\ref{eq:A_3x2x2}) does not have the full rank, i.e., all four 
\(3\times 3\) minors \(m_j\) in Eq.~(\ref{eq:A_3x2x2}) are zero.
The SLOCC orbits which appear inside \(X^{\vee}_{\rm sing}\) are essentially 
the same as the 3-qubit case.

Thus we obtain the partially ordered structure of multipartite entangled 
states as Fig.~\ref{fig:part_order}.
The tripartite entanglement consists of four classes.
Because the class of \(M\!-\! X^{\vee}\), whose representative is 
\(|000\rangle \!+\!|101\rangle \!+\!|110\rangle \!+\!|211\rangle\), 
and that of \(X^{\vee}\!-\! X^{\vee}_{\rm sing}\), whose representative is 
\(|000\rangle \!+\!|101\rangle \!+\!|211\rangle\), have the same local ranks 
\((3,2,2)\), they do not convert each other in the same reason as GHZ and W 
do not.  
However, the former two classes of the local ranks \((3,2,2)\) can convert to 
the latter two classes of \((2,2,2)\) by noninvertible local operations 
(i.e., LOCC). And we can "degrade" these tripartite entangled classes into 
the biseparable or completely separable classes by LOCC in a similar fashion 
to the \(3\) qubits.

We notice that 3 grades in the \(3\)-qubit case changed to 4 grades in 
the \(3 \!\times\! 2 \!\times\! 2\) (\(1\)-qutrit and \(2\)-qubit) case.
In general, the partially ordered structure becomes "higher",
as the system of each party becomes the higher dimensional one. 
We also see how the tensor rank \cite{note3} is inadequate for the 
onion-like classification of SLOCC orbits.

%%%%%%%%%%%%%%%%%%%%%%%%%%%%%%%%%%%%%%%%%%%%%%%%%%%%%%%%%%%%%%%%%%%%%%%%%%%%%%%
%
\subsection{\(n\!\geq\!4\)-qubit (format \(2^n\)) case}
\label{sec:nqubit}
Further in the \(n\!\geq\!4\)-qubit case, our classification works. 
The outermost class \(M(\!=\!\mathbb{C}P^{2^{n}\!-\!1})\!-\!X^{\vee}\) of 
generic \(n\)-partite entangled states is given by \({\rm Det}A_n \!\ne\! 0\). 
In \(n \!=\! 4\), \({\rm Det}A_4\) of degree 24 is explicitly calculated 
by the Schl\"afli's construction in Sec.~\ref{sec:schlafli}.
It would be suggestive to transform any generic \(4\)-partite state 
(\({\rm Det}A_4 \!\ne\! 0\)) to the "representative" of the outermost class 
by invertible local operations,
\begin{multline}
\label{eq:generic_4bit}
\alpha(|0000\rangle +|1111\rangle)
+\beta(|0011\rangle+|1100\rangle) \\
+\gamma(|0101\rangle +|1010\rangle)
+\delta(|0110\rangle+|1001\rangle),
\end{multline}
where the continuous complex coefficients \(\alpha,\beta,\gamma,\delta\) 
should satisfy
\begin{multline}
\label{eq:det4}
{\rm Det}A_4 = \alpha^2 \beta^2 \gamma^2 \delta^2 
(\alpha+\beta+\gamma+\delta)^2 (\alpha+\beta+\gamma-\delta)^2 \\ 
(\alpha+\beta-\gamma+\delta)^2 (\alpha-\beta+\gamma+\delta)^2 
(-\alpha+\beta+\gamma+\delta)^2 \\ \quad\:
(\alpha+\beta-\gamma-\delta)^2 (\alpha-\beta+\gamma-\delta)^2 
(\alpha-\beta-\gamma+\delta)^2 \ne 0.  
\end{multline}
Thus three complex parameters remain in the outermost class (since we 
consider rays rather than normalized state vectors).
This means that there are infinitely many same dimensional SLOCC orbits 
in the \(4\) qubits, and the SLOCC orbits never locally convert to each other
when their sets of the parameters are distinct.
It is also the case for the \(n \!>\! 4\) qubits.
Note that, in \(n\!=\!4\), this outermost class \(M\!-\!X^{\vee}\) corresponds
to the family of generic states in Verstraete {\it et al.}'s classification 
of the \(4\) qubits by a different approach (generalizing
the singular value decomposition in matrix analysis to complex 
orthogonal equivalence classes), and \(X^{\vee}\) contains their 
other special families \cite{verstraete+02}. 

The next outermost class is \(X^{\vee}\!-\!X^{\vee}_{\rm sing}\). 
In the \(4\) qubits, \(X^{\vee}_{\rm sing}\) is shown to consist of eight 
closed irreducible components of codimension 1 in \(X^{\vee}\); 
\(X^{\vee}_{\rm cusp},\: X^{\vee}_{\rm node}(\emptyset)\), and six 
\(X^{\vee}_{\rm node}(j_1,j_2)\) for \(1 \!\leq\! j_1 \!<\! j_2 \!\leq\! 4\) 
\cite{weyman+96}.
They neither contain nor are contained by each other.
Their intersections also give (finitely) many lower dimensional genuine
\(4\)-partite entangled classes. 
Since the \(4\)-partite entangled classes necessarily have the same local 
ranks \((2,2,2,2)\), these classes are not interconvertible by noninvertible 
local operations (i.e., any LOCC).
As typical examples, GHZ (the maximally entangled state in Bell's 
inequalities \cite{gisin+98}), 
\begin{equation}
|{\rm GHZ}\rangle = |0000\rangle + |1111\rangle,
\end{equation}
(i.e., \(a_{0000}=a_{1111} \ne 0\) and the others are \(0\)) is included in 
the intersection of \(X^{\vee}_{\rm node}(\emptyset)\) and six 
\(X^{\vee}_{\rm node}(j_1,j_2)\), but is excluded from \(X^{\vee}_{\rm cusp}\).
In contrast, W, 
\begin{equation}
|{\rm W}\rangle = |0001\rangle + |0010\rangle + |0100\rangle + |1000\rangle,
\end{equation}
(i.e., \(a_{0001}=a_{0010}=a_{0100}=a_{1000} \ne 0\) and the others are \(0\)) 
is included in the intersection of \(X^{\vee}_{\rm cusp}\) and  
six \(X^{\vee}_{\rm node}(j_1,j_2)\) but is excluded from 
\(X^{\vee}_{\rm node}(\emptyset)\).  

In the \(n\!>\!4\) qubits, \(X^{\vee}_{\rm sing}\) is shown to consist of 
just two closed irreducible components \(X^{\vee}_{\rm cusp}\) and 
\(X^{\vee}_{\rm node}(\emptyset)\) \cite{weyman+96}.
We find that GHZ and W are contained not only in 
\(X^{\vee}\:({\rm Det}A_n\!=\!0)\) but in \(X^{\vee}_{\rm sing}\); i.e.,
they have nontrivial solutions in Eqs.(\ref{eq:critical}), satisfying the
singular conditions.
They correspond to different intersections of further singularities, similarly 
to the \(4\) qubits. In other words, they are peculiar, living in the border 
dimensions between entangled states and separable ones, 

In brief, the dual variety \(X^{\vee}\) and its singularities lead to 
the {\it coarse} onion-like classification of SLOCC orbits, when SLOCC gives
infinitely many orbits.
The partially ordered structure of multipartite pure entangled states
becomes "wider", as the number \(n\) of parties increases.
Although many inequivalent \(n\)-partite entangled classes appear in the 
\(n\) qubits, they never locally convert to each other, as observed in 
\cite{dur+00}.
In particular, the majority of the \(n\)-partite entangled states never 
convert to GHZ (or W) by LOCC, and the opposite conversion is also
not possible. 
This is a significant difference from the bipartite or \(3\)-qubit case,
where almost any entangled state and the maximally entangled state 
(GHZ) can convert to each other by LOCC with nonvanishing probabilities.

%%%%%%%%%%%%%%%%%%%%%%%%%%%%%%%%%%%%%%%%%%%%%%%%%%%%%%%%%%%%%%%%%%%%%%%%%%%%%%%
%
\section{Conclusion}
\label{sec:conclude}
We have presented the onion-like classification of multipartite entanglement 
(SLOCC orbits) by the dual variety \(X^{\vee}\), i.e., the hyperdeterminant 
\({\rm Det}A\). 
It leads to the partially ordered structure, such as 
Fig.~\ref{fig:part_order}, of inequivalent multipartite entangled classes of 
pure states, which is significantly different from the totally ordered one 
in the bipartite case.
Local ranks are not enough to distinguish these classes, and we need to 
calculate SLOCC invariants associated with \({\rm Det}A\).
In other words, the generic entangled class of the maximal dimension 
(the outermost class) is given by the outside of \(X^{\vee}\) 
(\({\rm Det}A \ne 0\)), and other multipartite entangled classes 
appear as \(X^{\vee}\) or its different singularities.
Analytically, the classification of multipartite entanglement corresponds to
that of the number and pattern of the solutions in Eqs.~(\ref{eq:critical}).

This work reveals that the situation of the widely known bipartite or 
\(3\)-qubit cases, where the maximally entangled states in Bell's 
inequalities belong to the generic class, is exceptional.
Lying far inside the onion structure, the maximally entangled states (GHZ) 
are included in the lower dimensional peculiar class in general, e.g., for 
the \(n \geq 4\) qubits.
It suggests two points.
The majority of multipartite entangled states can not convert to GHZ by LOCC, 
and vice versa. So, we have given an alternative explanation to this 
observation, first made in \cite{dur+00} by comparing the number of local 
parameters accessible in SLOCC with the dimension of the whole Hilbert space.
Moreover, there seems no a priori reason why we choose GHZ states as
the {\it canonical} \(n\)-partite entangled states, which, for example,
constitute a minimal reversible entanglement generating set (MREGS) in
asymptotically reversible LOCC \cite{bennett+00,wu+00}.
Since the onion-like classification is given by every closed subset, 
not only our work enables us to see intuitively why, say in 
the \(3\) qubits, the W class is rare compared to the GHZ class, but 
it can be also extended to the classification of multipartite mixed states
(see Appendix~\ref{sec:mixed}).

The onion-like classification seems to be reasonable in the sense that
it coincides with the SLOCC classification when SLOCC gives finitely many 
orbits, such as the bipartite or \(3\)-qubit cases.
So two states belonging to the same class can convert each other by 
invertible local operations with nonzero probabilities.
On the other hand, when SLOCC gives infinitely many orbits, 
this classification is still SLOCC-invariant, but may contain in one class 
infinitely many {\it same} dimensional SLOCC orbits which can not locally 
convert to each other even probabilistically.
For example, in the \(4\)-qubit case, the generic entangled class in 
Eq.(\ref{eq:generic_4bit}) has three nonlocal continuous parameters.
Note that it can be possible to make the onion-like classification finer, by 
characterizing the nonlocal continuous parameters in each class.

Then, we may ask, what is the physical interpretation of the onion-like 
classification in the case of infinitely many SLOCC orbits? 
Although a simple answer has yet to be found, we discuss two points.
(i) Let us consider {\it global} unitary operations which create 
the multipartite entanglement.
On the one hand, states in distinct classes would have the different 
complexity of the global operations, since they have the distinct number 
and pattern of nonlocal parameters.
On the other hand, states in one class are supposed to have the equivalent
complexity, since they just correspond to different "angles" of the global 
unitary operations.
(ii) We can consider the case where {\it more than one} state are shared,
including the asymptotic case.
Even in two shared states, there can exist a local conversion which is 
impossible if they are operated separately, such as the catalysis 
effect \cite{jonathan+99}. 
So we can expect that we do it more efficiently in this situation, and 
the coarse classification may have some physical significance.
This problem remains unsettled even in the bipartite case. 

Finally, two related topics are discussed. 
(i) The absolute value \(|{\rm Det}A_n|\) of the hyperdeterminant, 
representing the amount of generic entanglement, is an entanglement monotone 
by Vidal \cite{vidal00}.
This never conflicts with the property that the maximally entangled states
in Bell's inequalities (GHZ) generally have a zero \({\rm Det}A_n\). 
A single entanglement monotone is insufficient to judge the LOCC 
convertibility, and generic entangled states of the nonzero \({\rm Det}A_n\) 
can not convert to GHZ in spite of decreasing \(|{\rm Det}A_n|\).
(ii) The \(3\)-tangle \(\tau = 4|{\rm Det}A_3|\) first appeared in the 
context of so-called entanglement sharing \cite{coffman+00}; i.e., 
in the \(3\) qubits, there is a constraint (trade-off) between the amount of 
\(2\)-partite entanglement and that of \(3\)-partite entanglement.
By using the entanglement measure (concurrence \(C\)) for the \(2\)-qubit 
{\it mixed} entangled states, this is written as 
\(C^2_{1(23)} \geq C^2_{12} + C^2_{13}\), and \(\tau\) is defined by 
\(\tau = C^2_{1(23)}-C^2_{12}-C^2_{13}\) for the \(3\)-qubit {\it pure} 
entangled states.
We expect that, in turn, the hyperdeterminant \({\rm Det}A_n\) gives a clue 
to find the entanglement measure of more than \(2\)-qubit {\it mixed} states.

\section*{Acknowledgments}
The author would like to thank M. Wadati, M. Murao, G. Kato, the members in 
the ERATO Project of Quantum Computation and Information, and the participants 
in the Sixth Quantum Information Technology Symposium in Japan on May (2002)
for the most helpful discussions. 
The work is partially supported by the ERATO Project.

%%%%%%%%%%%%%%%%%%%%%%%%%%%%%%%%%%%%%%%%%%%%%%%%%%%%%%%%%%%%%%%%%%%%%%%%%%%%%%%
%
%  appendices
%
\appendix
\section{Representatives of the \(3\)-qubit entangled classes}
\label{sec:rep_3qubit}

In Sec.~\ref{sec:4}, we have classified entangled classes (SLOCC orbits),
utilizing SLOCC-invariant closed subvarieties such as the dual variety
\(X^{\vee}\) and its singularities.
In this appendix, we give an intuitive explanation about our technique.
We obtain entangled classes {\it by their representatives}.
The \(3\)-qubit case is exemplified, and the notation and terminology of 
Sec.~\ref{sec:sing} is followed.
 
As the representative of the outermost generic entangled class, 
almost any state (indeed, satisfying \({\rm Det}A_3 \ne 0\)) 
in the whole space \(M={\mathbb C}P^7\) is qualified. 
The GHZ state \(|000\rangle + |111\rangle\) is chosen among them,
since it can be seen as the multidimensional analog of the identity matrix.

We look for the representative of the dual variety \(X^{\vee}\),
which is qualified as that of the next outermost entangled class.
When \(X^{\vee}\) is the hyperplane tangent to the Segre variety \(X\) at 
\(x^o\) such that \(x^{(j)}_{i_j}=\delta_{i_j,0}\) (\(j=1,2,3\)), 
the "\(x^o\)-section" of \(X^{\vee}\) is given as
\begin{equation}
\label{eq:xv_o}
X^{\vee}|_{x^o} = \{a_{000} = a_{001} = a_{010} = a_{100} =0 \},
\end{equation}
in order that Eqs.(\ref{eq:critical}) have the nontrivial solution \(x^o\).
This suggests that the representative of \(X^{\vee}\) is the W state 
\(|001\rangle + |010\rangle + |100\rangle\),
since the states given by Eq.(\ref{eq:xv_o}) and W convert to each other under 
some invertible local operations \(G \in GL_2 \times GL_2 \times GL_2\) as
\begin{align}
& a_{011}|011\rangle + a_{101}|101\rangle + a_{110}|110\rangle + 
a_{111}|111\rangle \nonumber \\
&\stackrel{G}{\sim}  |011\rangle + |101\rangle + |110\rangle + |111\rangle 
\nonumber \\
&\stackrel{G}{\sim}  |001\rangle + |010\rangle + |100\rangle.
\end{align}

The candidates for the next outer entangled class are two, node-type 
\(X^{\vee}_{\rm node}\) and cusp-type \(X^{\vee}_{\rm cusp}\), singularities
of \(X^{\vee}\).
We first consider the representative of \(X^{\vee}_{\rm node}(1)\).
According to Eq.(\ref{eq:node}), the \(x^o\)-section of 
\(X^{\vee}_{\rm node}(1)\) is given as
\begin{multline}
X^{\vee}_{\rm node}(1)|_{x^o} = X^{\vee}|_{x^o} \cap X^{\vee}|_{x^o(1)} \\
=\{a_{000}=a_{001}=a_{010}=a_{100}=a_{011}=a_{111}=0 \}.
\end{multline}
We find that the representative of \(X^{\vee}_{\rm node}(1)\) is the 
biseparable state \(|001\rangle + |010\rangle\) in \(B_1\), checking that
\begin{equation}
a_{101}|101\rangle + a_{110}|110\rangle 
\;\stackrel{G}{\sim}\; |001\rangle + |010\rangle.
\end{equation}
In the same manner, \(X^{\vee}_{\rm node}(2)\) or \(X^{\vee}_{\rm node}(3)\)
represents the biseparable class \(B_2\) or \(B_3\), respectively.

Let us second analyze the representative of \(X^{\vee}_{\rm cusp}\).
In terms of the quadric part \(y\) of \(F(A,x)\) at \(x^o\):
\begin{equation}
y = \left(\begin{array}{ccc}
    0 & a_{110} & a_{101} \\ a_{110} & 0 & a_{011} \\
    a_{101} & a_{011} & 0 
    \end{array}\right),
\end{equation}
the \(x^o\)-section of \(X^{\vee}_{\rm cusp}\) is given as
\begin{align}
X^{\vee}_{\rm cusp}|_{x^o} = \{& a_{000}=a_{001}=a_{010}=a_{100}=0, \nonumber\\
                               & \det y = 2 a_{011}a_{101}a_{110}=0 \}.
\end{align}
We have three possibilities for \(\det y =0\).
In the case of \(a_{011}=0\), this component of \(X^{\vee}_{\rm cusp}\)   
represents the biseparable class \(B_1\), since
\begin{equation}
a_{101}|101\rangle + a_{110}|110\rangle + a_{111}|111\rangle
\stackrel{G}{\sim} |001\rangle + |010\rangle.
\end{equation}
Likewise, in the case of \(a_{101}=0\) or \(a_{110}=0\), each component of
\(X^{\vee}_{\rm cusp}\) corresponds to the biseparable class \(B_2\) or 
\(B_3\), respectively.
Remembering that each \(B_j\) is characterized by \(X^{\vee}_{\rm node}(j)\) 
for \(j=1,2,3\), we have shown that \(X^{\vee}_{\rm cusp}\) has three 
irreducible components \(X^{\vee}_{\rm node}(j)\).
Thus, the next outer entangled classes are three biseparable classes \(B_j\),
which never contain nor are contained by each other.  

In general, remaining entangled classes are given by further singularities
of \(X^{\vee}\) such as combinations of the above \(X^{\vee}_{\rm node}\)
and \(X^{\vee}_{\rm cusp}\), or genuine higher singularities.
In the \(3\)-qubit case, since we see that  \(X^{\vee}_{\rm node}(j)\), 
representing the biseparable class \(B_j\), is just characterized as 
\({\mathbb C}P^{1}_{j{\rm \mbox{-}th}}\times {\mathbb C}P^{3}\),
there remains just one smaller closed irreducible subvariety 
\({\mathbb C}P^{1} \times {\mathbb C}P^{1} \times {\mathbb C}P^{1} 
= X\) as their intersection \(\bigcap_{j=1,2,3}X^{\vee}_{\rm node}(j)\).
This Segre variety \(X\) represents the completely separable class \(S\), 
whose representative is \(|000\rangle\).

In the text, we have carried out the above procedure in the "\(x^o\)-free" 
manner (\(x^o\) should be taken as {\it any} state on \(X\)), and have 
obtained entangled classes as (difference) subsets.
It enables us to decide readily which entangled class a given state 
\(|\Psi\rangle\) belongs to.
After the classification of entangled classes, we can clarify 
their partially ordered structure under noninvertible local operations
in the same manner as in the text.

\section{Classification of multipartite mixed states}
\label{sec:mixed}

The onion structure is also useful for the SLOCC-invariant classification of 
mixed entangled states.
A mixed state \(\rho\) can be written as a convex combination of projectors 
onto pure states (extremal points),
\begin{equation}
\label{eq:rho}
\rho = \sum_{\mu} \; p_{\mu} |\Psi_{\mu} ({\mathcal O}_{\lambda})\rangle 
       \langle\Psi_{\mu} ({\mathcal O}_{\lambda})|, 
\quad p_{\mu} > 0, 
\end{equation}
where \(|\Psi_{\mu}({\mathcal O}_{\lambda})\rangle\) is the pure 
state belonging to the SLOCC orbit \({\mathcal O}_{\lambda}\) of an index
\(\lambda\).
\(\lambda\) is labeled by the closed subvariety 
\(\overline{{\mathcal O}_{\lambda}}\) (i.e., the closure of 
\({\mathcal O}_{\lambda}\)) such as \(X^{\vee}\), \(X^{\vee}_{\rm node}(J)\), 
\(X^{\vee}_{\rm cusp}\), and \(X\).
Note that, in the multipartite case, there can be many closed subvarieties 
\(\overline{{\mathcal O}_{\lambda}}\) which never contain nor are contained 
by each other; for example, \(X^{\vee}_{\rm node}(1)\), 
\(X^{\vee}_{\rm node}(2)\), and \(X^{\vee}_{\rm node}(3)\) in 
Fig.~\ref{fig:onion_3bit}.
So, by taking the union of these "competitive" closed subvarieties
\(\overline{{\mathcal O}_{\lambda}}\) (it will form their convex hull 
in the space of \(\rho\)), we pick up only {\it totally ordered} ones, 
e.g., \(M\), \(X^{\vee}\), \(X^{\vee}_{\rm cusp}=\bigcup_{j=1,2,3} 
X^{\vee}_{\rm node}(j)\), and \(X\) in Fig.~\ref{fig:onion_3bit}, 
for convenience later.  
Now, we are concerned with at most how various classes of pure 
entangled states the mixed state \(\rho\) consists of. 
We take the maximal closure of \(\overline{{\mathcal O}_{\lambda}}\) appeared 
in Eq.(\ref{eq:rho}) and denote it by \(\overline{{\mathcal O}_{\rm max}}\).
However, since \(\rho\) can be decomposed into the form of Eq.(\ref{eq:rho}) 
in infinitely many ways, we should take the minimal closure of 
\(\overline{{\mathcal O}_{\rm max}}\) over all possible decompositions, and
write it as \(\min \overline{{\mathcal O}_{\rm max}}\). 
Every convex subset \({\mathcal S}_{\lambda}\) of 
\(\lambda = \min \overline{{\mathcal O}_{\rm max}}\) is closed such that 
\({\mathcal S}_{\lambda}\) of the smaller \(\lambda\) is contained by 
that of the larger one.
In other words, \({\mathcal S}_{\lambda}\) of the larger \(\lambda\) consists
of more classes (SLOCC orbits) of pure entangled states.
That is how the mixed state \(\rho\) is classified into the {\it closed} 
convex subsets \({\mathcal S}_{\lambda}\) under SLOCC.

In the bipartite case, \(\min \overline{{\mathcal O}_{\rm max}}\) is called 
the Schmidt number \cite{terhal+00} (since 
\(\overline{{\mathcal O}_{\lambda}}\) is just labeled by the Schmidt local 
rank, as seen in Sec.~\ref{sec:intro}). 
Also in the \(3\)-qubit case, this kind of the classification has been 
done in \cite{acin+01}, and four classes appear, following the 
above recipe: 
\medskip\\
(i) GHZ class \({\mathcal S}_{M} \!-\! {\mathcal S}_{X^{\vee}}\) (consisting 
of all pure states); \\ 
(ii) W class \({\mathcal S}_{X^{\vee}} \!-\! 
{\mathcal S}_{X^{\vee}_{\rm cusp}}\) (consisting of the pure
W, biseparable, or separable states); \\ 
(iii) biseparable class \({\mathcal S}_{X^{\vee}_{\rm cusp}} \!-\!
{\mathcal S}_{X}\) (consisting of the pure biseparable or separable states);\\ 
(iv) separable class \({\mathcal S}_{X}\) (consisting of only the pure 
separable states). 
\medskip

Needless to say, the trouble is considering all possible decompositions in
Eq.~(\ref{eq:rho}).
So, it is very difficult to give the criterion to distinguish each closed 
convex subset \({\mathcal S}_{\lambda}\), even to distinguish the separable 
subset \({\mathcal S}_{X}\).
Still, it would be interesting to observe that a witness operator 
\(\mathcal{W}\), which forms the tangent hyperplane 
\({\rm tr}(\rho\mathcal{W})\!=\!0\) detecting \({\mathcal S}_{\lambda}\) 
a given \(\rho\) belongs to, shares the same idea as our dual variety.

%%%%%%%%%%%%%%%%%%%%%%%%%%%%%%%%%%%%%%%%%%%%%%%%%%%%%%%%%%%%%%%%%%%%%%%%%%%%%%%
%
%  references
%


\begin{thebibliography}{~~}
\bibitem{note1}
SLOCC was tied to \(SL \times SL\) in \cite{dur+00}.
Since we treat \(|\Psi\rangle\) as a ray, i.e., an unnormalized vector, we 
rather relate SLOCC to \(GL \times GL\).
\bibitem{note2}
The local rank can be also defined as the rank of the reduced density matrix 
traced out for all except one party.
Note that this definition is applicable to the multipartite case.
\bibitem{note3}
The tensor rank means the number of terms in the minimal decomposition of 
\(|\Psi\rangle\) in Eq.(\ref{eq:multi}) by separable states.
It coincides with the local rank in the bipartite case.
\bibitem{note4}
For example, let us consider two Einstein-Podolsky-Rosen (EPR) pairs
\(\frac{1}{\sqrt{2}}(|00\rangle +|11\rangle)_{12}\otimes 
\frac{1}{\sqrt{2}}(|00\rangle + |11\rangle)_{34}\) in the \(4\) qubits.
Since their local ranks are \((2,2,2,2)\), we can not distinguish
this state from genuine \(4\)-partite entangled states 
(cf. Sec.~\ref{sec:nqubit}). In contrast, we readily find that the state is 
included in \({\mathbb C}P^{3} \!\times\! {\mathbb C}P^{3}\), so that it is 
separable (not genuine \(4\)-partite entangled).
\bibitem{note5}
If there exists a noninvertible local operation between SLOCC orbits, 
some of the local ranks decrease. 
Note that the converse is generally not the case.
However, in our examples in Sec.~\ref{sec:4}, we do find some noninvertible
local operations for the decrease of local ranks.
\end{thebibliography}
\end{document}